\documentstyle[12pt,epsfig]{article}

\def\be{\begin{equation}}
\def\ee{\end{equation}}
\def\ba{\begin{eqnarray}}
\def\ea{\end{eqnarray}}
\def\ga{\mathrel{\raise.3ex\hbox{$>$\kern-.75em\lower1ex\hbox{$\sim$}}}}
\def\la{\mathrel{\raise.3ex\hbox{$<$\kern-.75em\lower1ex\hbox{$\sim$}}}}

\begin{document}

\baselineskip=16pt 
\begin{titlepage} 
\rightline{CERN-TH/2002-221}
\rightline{UMN--TH--2112/02}
\rightline{TPI--MINN--02/41}
\rightline{hep-ph/0209036}
\rightline{September 2002}  
\begin{center}

\vspace{0.5cm}

\large {\bf Stable, Time-Dependent, Exact Solutions for \\ Brane Models
with a Bulk Scalar Field}
\vspace*{5mm}
\normalsize

{\bf  Panagiota Kanti$^{\,(a)}$, Seokcheon Lee$^{\,(b)}$} and 
{\bf Keith A. Olive$^{\,(b,c)}$}

\smallskip 
\medskip 

$^{\,(a)}${\it Theory Division, CERN, CH-1211 Geneva 23, Switzerland}
 
$^{\,(b)}${\it  School of Physics and
Astronomy,\\  University of Minnesota, Minneapolis, MN 55455, USA} 

$^{\,(c)}${\it Theoretical Physics Institute,
\\  University of Minnesota, Minneapolis, MN 55455, USA}

\smallskip 
\end{center} 
\vskip0.6in 
 
\centerline{\large\bf Abstract}

We derive two classes of brane-world solutions arising in the presence
of a bulk scalar field. For static field configurations, we adopt
a time-dependent, factorizable metric ansatz that allows for radion
stabilization. The solutions are characterized by a non-trivial
warping along the extra dimension, even in the case of a vanishing bulk
cosmological constant, and lead to a variety of inflationary,
time-dependent solutions
of the 3D scale factor on the brane. We also derive the 
constraints necessary for the stability of these solutions under
time-dependent perturbations of the radion field, and we demonstrate
the existence of phenomenologically interesting, stable solutions with
a positive cosmological constant on the brane.

\vspace*{2mm} 

\end{titlepage} 

\section{Introduction}
\setcounter{equation}{0}

Over the last few years, there has been considerable interest in models
in which our  universe is a 3-brane (a hyper-surface) embedded in a higher
dimensional bulk.  Much of the interest in extra-dimensional field theory
is due to the hope for a solution to the hierarchy problem
\cite{An,RS,hierarchy}. In these models, the extra dimensions  are hidden
from us, not necessarily by their smallness but by our confinement to
 a four-dimensional slice of the bulk spacetime \cite{An}. In 
contrast to Kaluza-Klein scenarios, standard model interactions are 
confined to a brane whereas gravity propagates through the bulk
perpendicular  to the brane.
The hierarchy problem can be resolved by either postulating large extra
dimensions (in which case the TeV scale is the fundamental scale of
gravity and the Planck scale is derived in terms of the fundamental 
scale and the volume of the extra-dimensional space) \cite{An}  or when
the 4D metric scales exponentially throughout the bulk (the so called
``warp" factor) \cite{RS}.

While static brane-world models have served as a useful tool for testing
ideas in higher dimensional spacetimes, their direct applicability to
cosmology is limited. More realistic cosmological models may be derived
by allowing a non-vanishing four-dimensional cosmological constant,
or by introducing time-dependent energy-densities on the branes. Various
cosmological aspects of such models have been investigated in the
literature \cite{large}-\cite{more2}. One of the serious problems in brane
models is the resulting unconventional set of Friedmann equations
\cite{LOW, LK, BDL}. The Hubble parameter, $H$, on the brane is often
found to scale as $H \sim \rho$, rather than the standard four dimensional
dependence, $H \sim \sqrt{\rho}$. It has been shown however,
that this problem can be solved upon the proper stabilization of the
extra dimension \cite{kkop1}-\cite{kkop2}, that removes any unnecessary
constraints between the brane energy-densities.  

Both static and time-dependent generalizations of the original
Randall-Sundrum solutions \cite{RS} have been also constructed by
introducing a bulk scalar field \cite{scalars, kop1}. As a matter of
fact, the task of the stabilization of the extra dimension was first
accomplished by introducing a bulk scalar field, which had different
non-vanishing vacuum expectation values on each of the two branes
\cite{GW}. The same topic was further elaborated in \cite{kop1,stab-scal}.
Here, we will derive two classes of brane-world solutions which include
a static bulk scalar field and provide inflationary solutions in the 4D
slices. The two classes correspond to either vanishing or non-vanishing
bulk potential for the scalar field, and are characterized by a non-trivial 
warping of the metric along the extra dimension even in the case of zero
bulk cosmological constant. Due to the appearance of a bulk curvature
singularity, we are forced to consider two-brane-system configurations
which, however, have a fixed inter-brane distance and lead to 
conventional FRW equations on the branes without any additional
fine-tuning. By using a method developed recently in Ref. \cite{kop},
we study the stability of those solutions under time-dependent
perturbations of the radion field, and demonstrate that we can
easily find parameter regimes where these solutions are {\em
stable}. Even more important is the fact that some of these stable
solutions have a {\it positive} cosmological constant on the
brane - previously all known solutions of this type were unstable
\cite{dS}.

We organize this paper as follows. In section 2, we present the
equations of motion of our theory and show how a factorizable (in time
and the extra space coordinate $y$) scale factor can be obtained in 
the presence of a bulk scalar field. In section 3, we present two
classes of inflationary brane solutions involving a static bulk field
with vanishing or non-vanishing, respectively, bulk potential and
we demonstrate that these solutions indeed lead to conventional FRW
equations. We derive the conditions for the stabilization of these
solutions in section 4 and investigate the parameter regimes that 
correspond to stable configurations.  Finally, in section 5, we
summarize our results.

\section{Equations of Motion for Gravity and a Bulk Scalar Field}
\setcounter{equation}{0}

We start from the cosmological principle of isotropy and homogeneity in the 
three space-like dimensions of the brane. The presence of the brane breaks the 
isotropy along the fifth dimension and this is reflected in the explicit
$y$-dependence of the metric tensor. Based on these facts, we make the
following ansatz:
 \be
  ds^2 = g_{MN}\,dx^M dx^N=
  -n^2(t,y)\,dt^2 + a^2(t,y)\,\gamma_{ij}\,dx^i dx^j + b^2(t,y)\,dy^2\,,
  \label{metric}
 \ee
where $M,N=0,1,2,3,5$, $\gamma_{ij}$ is the usual Robertson-Walker 3-space
metric tensor, and $t$, $x^i$ ($i=1,2,3$), and $y$ are the time- and 
space-like coordinates along the brane and the extra dimension, respectively.

In addition, we consider a bulk scalar field $\phi(t,y)$, which depends only
on time and the extra coordinate\footnote{Again, we
use the standard assumption that the 3-dimensional space is homogeneous
and isotropic. Thus, the field is independent of the 3-dimensional spatial
variables.}. The action of this five-dimensional, gravitational theory
is given by:
\ba
&~& \hspace*{-2.5cm} S=-\int d^{4}x\,dy\,\sqrt{-g}\,\Biggl\{-\frac{
M_{5}^{3}}{16\pi}\,\hat{R}+\Lambda _{B}+\frac{1}{2}\,\partial _{M}\phi
\,\partial ^{M}\phi+V_{B}(\phi) \nonumber \\
&~& \hspace{5cm}+ \sum_i\,\Bigl[\Lambda _i+V_i(\phi)\Bigr]\,
\frac{\delta (y-y_i)}{b} \Biggr\}\,,
\label{action}
\ea
where $M_{5}$ is the fundamental, five-dimensional Planck mass, $\hat{R}$
denotes the five-dimensional scalar curvature, and $V_B$ and $V_i$ stand for
the bulk and brane potentials, respectively, of the scalar field. Finally,
$\Lambda_{B}$ and $\Lambda_i$ are the vacuum energies of the bulk and the
branes. From  Eq. (\ref{action}), one can derive the scalar field equation
of motion:
\smallskip
\be
\frac{1}{n^{2}}\ddot{\phi}-\frac{1}{b^{2}}\phi''-\frac{1}{n^{2}}\,
\Biggl(\frac{\dot{n}}{n}-3\frac{\dot{a}}{a}-\frac{\dot{b}}{b}\Biggl)\dot{\phi}
-\frac{1}{b^{2}}\Biggl(\frac{n'}{n}+3\frac{a'}{a}-\frac{b'}{b}\Biggl)\phi'
+ \frac{\partial V_{B}}{\partial \phi} 
+\sum_i \frac{\partial V_i} {\partial\phi}\,\frac{\delta (y-y_i)}{b} = 0\,,
\label{fieldeq}
\ee
where dots and primes denote derivatives with respect to $t$ and $y$,
respectively.

The matter content of the five-dimensional space-time is described by the
energy-momentum tensor of the bulk scalar field and the bulk cosmological
constant, which may be written as:
\be
T^{M}_{N} = -\Lambda_{B}\,\delta^{M}_{N} 
-\sum_i (V_i + \Lambda_i)\,\,\frac{\delta (y-y_i)}{b}\,
\delta^{M}_{\mu}\,\delta^{\nu}_{N}\,\delta^{\mu}_{\nu}
                  + T^{M}_{N}(\phi)\,,
\ee		     
where $\mu, \nu$ = 0, 1, 2, 3,  and
\be
T_{MN}(\phi) = \partial_M\phi\,\partial_N\phi - 
g_{MN}\,\Biggl[\frac{1}{2}\,\partial_P\phi\,\partial^{P}\phi + 
V_{B}(\phi)\Biggr]\,.
\ee

We next consider the five dimensional set of Einstein equations. They can be
obtained by the variation of the action (\ref{action}) with respect to the
metric, and, for the metric ansatz (\ref{metric}), they take the form
\footnote{Our convention for the Riemann curvature tensor is 
 $R^{\mu}{}_{\nu\rho\lambda}
= \partial_\rho \Gamma^\mu_{\nu\lambda} - \partial_\lambda \Gamma^\mu_{\nu\rho} 
+ \Gamma^\eta_{\nu\lambda}\Gamma^\mu_{\rho\eta} -
\Gamma^\eta_{\nu\rho}\Gamma^\mu_{\lambda\eta}$.} 
\ba
G_{00} &=& 3\Biggl\{ -\frac{n^2}{b^2}\,\Biggl[\frac{a''}{a} +
\frac{a'}{a}\,\Biggl(\frac{a'}{a} - \frac{b'}{b}\Biggr)\Biggr] 
+\frac{\dot{a}}{a}\,\Biggl(\frac{\dot{a}}{a} +
\frac{\dot{b}}{b}\Biggr) +k\frac{n^2}{a^2}\Biggr\}
= \kappa_5^2 \, T_{00}\,,\label{00}\\[3mm] 
G_{ii} &=& \frac{a^2}{b^2}\,\gamma_{ii}\Biggl\{\frac{a'}{a}\,
\Biggl(\frac{a'}{a} + 2\frac{n'}{n}\Biggr) -\frac{b'}{b}\,\Biggl(\frac{n'}{n}
+2\frac{a'}{a}\Biggr) +2 \frac{a''}{a} +\frac{n''}{n}\Biggr\}\nonumber\\[2mm]
&+&\frac{a^2}{n^2}\,\gamma_{ii}\Biggl\{-2\frac{\ddot{a}}{a} 
+ \frac{\dot{a}}{a}\,\Biggl(-\frac{\dot{a}}{a} +
2\frac{\dot{n}}{n}\Biggr) -\frac{\ddot{b}}{b} + \frac{\dot{b}}{b}\,
\Biggl(-2\frac{\dot{a}}{a} + \frac{\dot{n}}{n}\Biggr) \Biggr\} - k \gamma_{ii}
=\kappa_5^2\,T_{ii}\,,
\label{ii}\\[3mm]
G_{05} &=& 3\Biggl(\frac{n'}{n} \frac{\dot{a}}{a}
+ \frac{a'}{a} \frac{\dot{b}}{b} -\frac{\dot{a}'}{a}\Biggr)= 
\kappa_5^2\,T_{05}\,,
\label{05}\\[3mm]
G_{55} &=& 3\Biggl\{\frac{a'}{a}\,\Biggl(\frac{a'}{a} +
\frac{n'}{n}\Biggr) -\frac{b^2}{n^2}\,\Biggl[\frac{\dot{a}}{a}\,
\Biggl(\frac{\dot{a}}{a}-\frac{\dot{n}}{n}\Biggr) +
\frac{\ddot{a}}{a}\Biggr]-k\frac{b^2}{a^2}\Biggr\} = \kappa_5^2\,
T_{55}\,, 
\label{55}
\ea 
where $\kappa_5^2=8\pi G_5=8\pi/M_5^3$ is the five-dimensional Newton's
constant, and $k=0, \pm 1$ denotes the spatial curvature of the
four-dimensional spacetime along the brane.

While the metric is continuous across the extra dimension, its derivatives
with respect to $y$ can be discontinuous because of the inhomogeneity of the
matter distribution in the fifth dimension, notably the branes. Therefore, a
delta function appears in the Einstein tensor and this must be matched with
a delta function in the energy-momentum tensor \cite{I}. Here we use a
similar notation for the {\it jump} of the scale factors\footnote{Here $a''= 
a_{R}'' + [a']_i\,\delta(y-y_i)$ where $a_{R}''$ is the non-distributional part
of the second derivative of $a$, and $[a']_i$ is the {\it jump} in the
first derivative across $y = y_i$, defined by $[a']_i = a'(t,y=y_i+\epsilon) -
a'(t,y=y_i-\epsilon)$.} as in references \cite{BDL,kkop1} and obtain
\ba
&~&\frac{1}{b_{i}}\frac{[a^{\prime }]_i}{a_{i}} = -\frac{\kappa_5^{2}}{3}\,\Bigl 
[\Lambda _i+V_i(\phi_{i})\Bigr]\,, 
\label{jumpa}\\[2mm]
&~&\frac{1}{b_{i}}\frac{[n^{\prime }]_i}{n_{i}} = -\frac{\kappa_5^{2}}{3}\,\Bigl 
[\Lambda _{i}+V_{i}(\phi_{i})\Bigr]\,,
\label{jumpn}
\ea
where the subscript $i$ denotes evaluation of all quantities at $y= y_i$.
The corresponding {\it jump} condition for the scalar field 
$\phi$ is obtained from the field equation (\ref{fieldeq})
\be
\frac{1}{b_{i}}\,[\phi^{\prime}]_i = 
\frac{\partial V_{i}(\phi)}{\partial \phi} \Biggl|_{y=y_i}\,. 
\label{jumpphi}
\ee
While the {\it jump} conditions for the scale factors (\ref{jumpa}) and
(\ref{jumpn}) are non-trivial given any inhomogeneous source on the
brane, the {\it jump} condition for the bulk field (\ref{jumpphi}) depends 
only on the interaction between the bulk field and the brane.  In particular,
if the  bulk field is sitting at an extremum on the brane, then the {\it jump}
of the scalar field (\ref{jumpphi}) is zero.

The facility to find an exact solution to the field equations
is often aided by our ability to factorize the scale
factor on the brane. Consider the following, 
\be
\frac{d}{dt}\frac{[a^{\prime}]_i}{a_{i}} = -\frac{\kappa_5^2}{3}\,b_{i}
\,\frac{\partial V_{i}}{\partial \phi}\,\frac{d\phi}{dt}\Biggl |_{y=y_i} 
- \frac{\kappa_5^2}{3}\,\dot{b}_{i}\,[\Lambda_{i} + V_{i}(\phi_{i})]\,.
\label{factorizable}
\ee 
If the scale factor can be factorized, then the right-hand-side of 
Eq. (\ref{factorizable}) must vanish. From Eq. (\ref{factorizable}), we
note the following:  if the field is sitting at a local minimum on the
brane {\em or} the field is static on the brane, our ability to factorize
the scale factor into independent $t$ and $y$ dependencies is tied to
the stability of the extra dimension on the brane ($\dot b_i =0$).
Off the brane, a corresponding relation can be found by examining the
$G_{05}$ component of the Einstein equation (\ref{05}), which can be
rewritten as
\be
\Biggl( \frac{n'}{n} - \frac{a'}{a}\Biggr) \frac{\dot{a}}{a} + \frac{a'}{a}
\frac{\dot{b}}{b} - \frac{d}{dt} \Biggl(\frac{a'}{a}\Biggr) =
\frac{\kappa_5^2}{3}\,\dot{\phi} \phi'\,.
\label{05'}
\ee
Without loss of generality we can write $n(t,y) = n(y)$. Then, Eqs.
(\ref{jumpa})-(\ref{jumpn}) lead to the following factorizable form 
of the 3D scale factor \footnote{A static four-dimensional universe can
be obtained by $n(t,y) = a(t,y) = n(y)$. This was the ansatz used in
Ref. \cite{RS} to solve the hierarchy problem and obtain conventional
Newtonian gravity.} 
\be
a(t,y) = a(t)\,n(y)\,.
\label{conformal2}
\ee
Inserting the above into Eq. (\ref{05'}) shows us that
\be
\frac{n'}{n}\frac{\dot{b}}{b} = \frac{\kappa_5^2}{3}\,\dot{\phi}\,\phi'\,.
\label{stationary}
\ee
Thus, if the bulk field is either time-independent or $y$-independent,
our condition for the factorization of the scale factor is intimately
connected to the stability of the fifth coordinate. In what follows,
we are going to assume that $\dot \phi=\dot b=0$, which brings our metric
ansatz to the form
\be
ds^2 = n^2(y)\,\Bigl[ -dt^2 + a^2(t)\,\gamma_{ij}\,dx^i dx^j \Bigr] 
+ b^2(y)\,dy^2\,.
\label{metric1}
\ee

\section{Exact Time-Dependent Solutions}
\setcounter{equation}{0}

We now proceed to derive exact solutions of the coupled system of 
gravitational and scalar field equations that have a non-static,
four-dimensional line-element. We are going to present two classes
of solutions: the first class corresponds to a vanishing potential,
$V_B(\phi)$, for the bulk scalar field, while the second one 
arises in the presence of a non-trivial (exponential, in a certain
limit) bulk potential.

\subsection{Solutions with a vanishing bulk potential: $V_{B} = 0$}

In our first example, the assumption of $V_{B}(\phi) = 0$ leads to a
simplification of the field equations in the bulk. The derivation
of the solution is further facilitated by a transformation of the
$y$-coordinate
that allows us to write the metric ansatz (\ref{metric1}) in terms of
``conformal" coordinates, i.e. $b(y) = n(y)$. Then, Einstein's equations
may be rewritten as (henceforth, $a$ denotes the 3D scale factor
appearing in the metric ansatz (\ref{metric1})): 
\ba
&& -\frac{3 n''}{n} +3\biggl(\frac{\dot{a}^2}{a^2} +\frac{k}{a^2}\biggr)
= \kappa_5^2\,\biggl( {\phi'^2 \over 2} + n^2
\Lambda_B \biggr)\,,\label{100}\\[3mm] 
&&
\frac{3 n''}{n} -\biggl(\frac{2\ddot{a}}{a} 
+\frac{\dot{a}^2}{a^2} + \frac{k}{a^2}\biggr) 
=- \kappa_5^2\,\biggl( {\phi'^2 \over 2} + n^2
\Lambda_B \biggr)\,,
\label{1ii}\\[4mm]
&& \frac{6 n'^2}{n^2}-3\biggl(\frac{\dot{a}^2}{a^2} +
\frac{\ddot{a}}{a}+\frac{k}{a^2}\biggr) = \kappa_5^2\,
\biggl({\phi'^2 \over 2} - n^2 \Lambda_B \biggr)\,.
\label{155}
\ea 
By combining Eqs. (\ref{100}) and (\ref{1ii}) we obtain
\be
\frac{\ddot{a}}{a} - \Biggl(\frac{\dot{a}}{a}\Biggr)^2 -\frac{k}{a^2} =0\,,
\label{a}
\ee
from which we can find the following time-dependent, inflationary solutions
for the scale factor~\footnote{In order to address the graceful exit
problem, we would be required to consider a non-static bulk field
which is beyond the scope of this paper.}
\be
a(t) = \left\{ \begin{array}{ll}
                  e^{H (t-t_0)} & \mbox{when $k=0$} \\[2mm]          
                  {1 \over H}\,\sinh[H (t-t_0)]
		   & \mbox{when $k=-1$} \\[2mm]
                  \frac{1}{H}\,\cosh[H (t-t_0)]  
	        & \mbox{when $k=+1$}
		 \end{array}
       \right. 
\label{inflation}
\ee
where $H$ and $t_0$ are constants. For $k=0$, we may obtain Minkowski, de
Sitter or Anti de Sitter solutions on the brane when $H^2$ has either a zero,
positive or negative value, respectively. For $k=-1$, we may have solutions
for either positive or negative values of $H^2$ (in the latter case, the
sinh-like solution above is replaced by $\sin[\tilde H (t-t_0)]/\tilde H$,
with $\tilde H^2=-H^2 >0$) while for $k=+1$ only solutions with $H^2>0$ are
allowed.

Turning to the scalar field equation of motion (\ref{fieldeq}), we see
that it now becomes
\be
\phi''+ 3\Biggl(\frac{n'}{n} \Biggr) \phi' = 0\,.
\label{fieldeq2}
\ee
The above equation can be integrated once to give
\be
\phi'(y) = \frac{c}{n^3(y)}\,,
\label{backreaction2}
\ee
where $c$ is an integration constant. The relation between $\phi'$
and $n$, derived above, can be used in Einstein's equations to derive the
form of the warp factor. In addition, we note that all solutions for
the scale factor $a(t)$ appearing in Eq. (\ref{inflation}) satisfy
the relation
\be
\frac{\ddot{a}}{a} =
\Biggl( \frac{\dot{a}}{a} \Biggr)^2 + \frac{k}{a^2} = H^2\,.
\label{k-ind}
\ee
Inserting both (\ref{backreaction2}) and (\ref{k-ind}) into the 
simplified set of field equations (\ref{100})-(\ref{155}), we
obtain 
\ba
\frac{n''}{n} &=& H^2 -\frac{\kappa_5^2}{3} \Biggl(\frac{c^2}{2 n^6}
+ n^2 \Lambda_{B} \Biggr)\,, 
\label{stationary003} \\[1mm]
2 \Biggl(\frac{n'}{n} \Biggr)^2 &=& 2 H^2 -\frac{\kappa_5^2}{3}
\Biggl(-\frac{c^2}{2 n^6} + n^2 \Lambda_{B} \Biggr)\,. 
\label{stationary553}
\ea

The above equations can be solved if we set $\Lambda_{B} = 0$. In that case,
we find the following solution:
\be
n^3(y) = \frac{\sinh(3 H |y|)}{\sinh(3 H |y_1|)}\,, 
\qquad {\rm where} \quad H^2=\frac{\kappa^2_5}{12}\,c^2\,\sinh^2(3H |y_1|)\,,
\label{n=b}
\ee
for a four-dimensional de Sitter spacetime ($H^2>0$). The corresponding
solutions for Min\-kowski and Anti de Sitter 4D spacetime are straightforward
to derive by taking the limits $H \rightarrow 0$ and $H \rightarrow i \tilde H$,
respectively. In the former case, the warp factor is linear in $y$, while,
in the latter, the solution is given in terms of a sin-like function. For
simplicity,  we will, for the remainder of this section, concentrate on
the case with $H^2>0$. It is worth noting that, although we have assumed a
vanishing bulk cosmological constant, non-trivial warping arises in all
cases where the expansion rate,  $H$, on the brane is non-zero. As can
be seen from the second of Eqs. (\ref{n=b}), the expansion rate is
closely related to the kinetic term of the scalar field in the bulk
which, in a way, replaces the bulk cosmological constant.

The Ricci scalar of the five-dimensional spacetime described by the metric
ansatz (\ref{metric1}) has the form :
\ba
\hat R &=& \frac{1}{n^2}\,\biggl(\frac{6\dot{a}^2}{a^2} + 
\frac{6\ddot{a}}{a} + \frac{6k}{a^2}\biggl) -
\biggl(\frac{4 n'^2}{n^4}+\frac{8n''}{n^3}\biggl) \label{Ricci1}\\[1mm]
&=&  \frac{\kappa_5^2}{n^2} \phi'^2\,. \nonumber 
\ea
The second equality can be obtained by using Eqs. (\ref{1ii})-(\ref{155}),
together with (\ref{a}), with $\Lambda_B=0$, in Eq. (\ref{Ricci1}).
As we can see, there is a true singularity in the bulk, at $y=0$. If we
choose to place the first brane at $y=y_1<0$ (in which case, we have the
normalization $n(y_1)=1$), then, in order to remove this singularity,
a second brane must be introduced at a point $y=y_2$, with $y_1<y_2<0$, so
that the singular point is never encountered.

We next examine the {\it jump} conditions imposed on the warp factor and
the scalar field at the positions of the two branes at $y_{1}$ and $y_{2}$,
respectively. Starting from Eqs. (\ref{jumpn}) and (\ref{jumpphi}), 
using the `conformal gauge' $b(y)=n(y)$, and inserting the solutions found
above, these conditions can be rewritten as
\ba
V_1 + \Lambda_1=\frac{6H}{\kappa^2_5}\,\coth(3 H |y_1|)\,, &~&  \quad
2c =\frac{\partial V_{1}(\phi)}{\partial \phi} \Biggl|_{y=y_{1}} 
\,,\label{jump1}\\[2mm]
V_2 + \Lambda_2=-\frac{6 H}{\kappa^2_5\,n_2}\,\coth(3 H |y_2|)\,, &~& 
\quad
\frac{2c}{n_2^3}=
-n_2\,\frac{\partial V_{2}(\phi)}{\partial \phi} \Biggl|_{y=y_{2}}\,,
\label{jump2}
\ea
where we have used the normalization condition $n_1=1$. As in the case of
the two-brane static solution with a bulk scalar field found in Ref. 
\cite{kop1}, the form of the interaction terms, $V_i$, completely
determines the ratio of the warp factors evaluated on the branes
\be
\Bigl( \frac{n_{1}}{n_{2}} \Bigr)^4 = \Biggl(
\frac{\sinh(3 H y_1)}{\sinh(3 H y_2)}\Biggr)^{\frac{4}{3}} = 
- \frac{(\partial_{\phi} V_{2})_{y = y_{2}}}
{(\partial_{\phi} V_{1})_{y = y_{1}}}\,.
\label{n0/nl}
\ee
Here too, the derivatives of the interactions are required to have
opposite signs, just like the total self-energies of the two branes.
Last, but not least, we may also observe that the inter-brane distance
can be derived from the above relation which is accompanied, through
the warp factor {\it jump} conditions, by the fine-tuning of only
one of the two brane self-energies. 

Let us briefly comment on the  relation between the expansion
rate $H$ and the brane potentials $V_i$. To do so, we must first define
the four-dimensional Newton's constant. Recalling the gravitational
part of our original action (\ref{action}) and noting that the first
combination inside brackets in Eq. (\ref{Ricci1}) stands for the 4D
scalar curvature, we may write
\ba \frac{1}{2 \kappa_{5}^2} \int d^4x\,dy\,\sqrt{-g}\,\hat R &=&
\frac{1}{2 \kappa_{5}^2} \int d^4x\,dy\,\sqrt{-g_{4}}\,n^5\,
\Bigl(\frac{1}{n^2}\,{\cal R}^{(4)} + \cdots \Bigr) \nonumber \\
&\equiv& \frac{1}{2 \kappa_{4}^2}
\int d^4x\,\sqrt{- g_{4}}\,( {\cal R}^{(4)} + \cdots )\,, 
\label{effective4}
\ea
where $\sqrt{- g_{4}} = a^3(t)$.  Thus, we have
\be \frac{1}{\kappa_{4}^2} \equiv \frac{1}{\kappa_{5}^2}
\oint dy\,n^3 = \frac{2}{\kappa_{5}^2}
\int_{y_{1}}^{y_{2}} dy\,n^3 =  - \frac{2}{\kappa_{5}^2}\,\frac{n_{i}^3}{3
H}\,\coth(3H|y_{i}|) 
\Biggl|_{y_{1}}^{y_{2}} = 
\sum_{i=1}^{2} n_{i}^4\,\frac{V_{i} + \Lambda_{i}}{9 H^2}\,, 
\label{Newton1} \ee 
where we have used the jump conditions (\ref{jump1}) and (\ref{jump2}) to
obtain the last equality. Note that although it appears that $\kappa_4$ is
independent of $\kappa_5$, $(V_i + \Lambda_i)$ and $H$ are interdependent
through $\kappa_5$. 

Finally, to check the Hubble equation on the brane, we must compute the
effective cosmological constant, $\Lambda_{eff}$, defined as
\be
\Lambda_{eff} =\oint dy\,n^5\,
\Biggl\{- \frac{\hat R^{(y)}}{2 \kappa^2_5}  
+ {\cal L} \Biggr\}\,,
\label{eff1}
\ee 
where $\hat R^{(y)}$ is given by the second two terms of Eq.
(\ref{Ricci1}) and $\cal L$ contains all of the non-gravitational
pieces of (\ref{action}). The result of the integration (taking care to
eliminate the boundary terms associated with $n''$) is
\be
\Lambda_{eff} =  \sum_{i=1}^{2} n_{i}^4\,\frac{V_{i} +\Lambda_{i}}{3}\,, 
\label{eff2}
\ee 
which when combined with Eq. (\ref{Newton1}), leads to the standard form
of the Friedmann equation,
\be
H^2 = \frac{\kappa_4^2}{3}\,\Lambda_{eff}
\label{Friedmann1}
\ee
without the need of any fine-tunings.

We will return to this solution in section 4, to test its stability under
small time-dependent perturbations.


\subsection{Solutions with a non-trivial bulk potential: $V_{B} \neq 0$}

In this subsection, we allow for a non-zero bulk potential, $V_{B} \neq
0$, a fact which will modify the equations in the bulk and subsequently
their solution. Although we retain the form of the metric (\ref{metric1}),
we choose to work with non-conformal coordinates; therefore, we redefine
the $y$-coordinate and fix the scale factor $b$ to a constant value,
i.e. $b=1$. Under the above assumptions, Einstein's equations become
\ba
&~&  -3n^2\,\biggl(\,\frac{n''}{n} + \frac{n'^2}{n^2}\biggr) + 
3\biggl(\frac{\dot{a}^2}{a^2} +\frac{k}{a^2}\biggr) = 
 \kappa_5^2 \, n^2\biggl( \frac{\phi'^2}{2} 
+ V_{B} + \Lambda_{B} \biggr)\,,
\label{00'}\\[3mm]
&~& 3n^2\,\biggl(\,\frac{n''}{n} + \frac{n'^2}{n^2}\biggr)
-\biggl(\frac{2\ddot{a}}{a} 
+\frac{\dot{a}^2}{a^2} + \frac{k}{a^2}\biggr) 
= -\kappa_5^2\,n^2 \biggl( \frac{\phi'^2}{2} + V_{B} +
\Lambda_{B} \biggr)\,,
\label{ii'}\\[3mm]
&~& 6 n^2\,\frac{n'^2}{n^2} -3\biggl(\frac{\dot{a}^2}{a^2}
+\frac{\ddot{a}}{a}+\frac{k}{a^2}\biggr) = 
\kappa_5^2\,n^2\,\biggl(\frac{\phi'^2}{2}-V_{B}-\Lambda_{B}\biggr)\,. 
\label{55'}
\ea   

By combining Eqs. (\ref{00'}) and (\ref{ii'}), we can once again derive
Eq. (\ref{a}) indicating that our previous solutions for the
scale factor $a(t)$, appearing in Eq. (\ref{inflation}), are still valid.
What remains to be found is the new form of the warp factor and the
scalar field. Plugging in the solutions for $a(t)$, the above
gravitational equations become
\ba
\frac{n''}{n} + \frac{n'^2}{n^2} &=& \frac{H^2}{n^2}
-\frac{\kappa_5^2}{3}\,\biggl(\frac{\phi'^2}{2} + V_{B}(\phi)
+\Lambda_{B} \biggr)\,,
\label{stationary00}\\[2mm]
\frac{2n'^2}{n^2} &=& \frac{2 H^2}{n^2} 
-\frac{\kappa_5^2}{3}\,\biggl(-\frac{\phi'^2}{2} + 
V_{B}(\phi) + \Lambda_{B} \biggr)\,.
\label{stationary55}
\ea
For $H=V_B=0$, the above equations lead to the static solutions with a bulk
scalar field found in Ref. \cite{kop1}. In general, Eqs. 
(\ref{stationary00})-(\ref{stationary55}) are difficult to solve,
nevertheless, we can find a sub-class of solutions which satisfy the
following:
\be
\frac{\phi'^2}{2} + \frac{1}{3}\,V_{B}(\phi) = E\,,
\label{case2}
\ee
where $E$ is a constant. The above relation can be used to provide the
solution for both the scalar field and the warp factor. In the bulk, the
scalar field equation of motion has the form
\be
\phi''+ 4\Biggl(\frac{n'}{n} \Biggr) \phi' = {dV_B \over d\phi}\,.
\label{fieldeqb}
\ee
By differentiating Eq. (\ref{case2}) with respect to $y$, we get 
\be
\phi'\,\phi'' + \frac{1}{3}\,{d V_{B} \over d \phi}\,\phi' = 0\,,
\ee
which, when combined with Eq. (\ref{fieldeqb}), yields the simple equation 
\be
\phi'' + \frac{n'}{n} \phi' = 0\,,
\label{fieldeq3}
\ee
whose solution is
\be
\phi'(y) = \frac{c}{n(y)}\,,
\label{backreaction3'}
\ee 
where $c$ is again an integration constant.

Going back to Eqs. (\ref{stationary00})-(\ref{stationary55}), we can now
combine them to get a single differential equation for $n(y)$
\be
\frac{n''}{n} = - \frac{\kappa_5^2}{6}\,(3E + \Lambda_B)\,,
\label{single}
\ee
leading to the following solution for the warp factor 
\be
n(y) = \left\{ \begin{array}{cl}
                 \frac{|y|}{|y_{1}|}, & \quad
		 \mbox{where \,$\chi^2 y_{1}^2 = 1$}\,, \\[2mm]
		 \frac{\sin(\omega |y|)}{\sin(\omega |y_1|)}, &
	\quad \mbox{where $\chi^2 \sin^2(\omega y_1) = \omega^2$}\,, \\[2mm]
		 \frac{\sinh(\omega |y|)}{\sinh(\omega |y_1|)}, &
	\quad \mbox{where $\chi^2 \sinh^2(\omega y_1) = \omega^2$}\,.
	       \end{array}
	\right.
\label{sol-n}		 
\ee
In the above, we have imposed the normalization condition $n_1=n(y_1)=1$,
and defined
\be
\omega^2 = -\frac{\kappa_5^2}{6}\,(3 E + \Lambda_{B})
\label{omega2}
\ee 
and
\be
\chi^2 = H^2 + \frac{c^2 \kappa_5^2}{3}\,.
\label{xi}
\ee
The three solutions presented in Eq. (\ref{sol-n}) correspond to the
combination $(3 E + \Lambda_{B})$ being zero, positive or negative,
respectively. All three solutions are characterized by a true
singularity at $y=0$. If we place, as in the previous section, a brane
at $y=y_1<0$, then, a second brane should be introduced at $y=y_2$,
with $y_1<y_2<0$,  if we want the singular point shielded.

The bulk potential of the scalar field is defined through Eq.
(\ref{case2}). The exact form can be easily derived by using the
expression of $\phi'$, in terms of $n(y)$, according to Eq. 
(\ref{backreaction3'}). The potential is everywhere well defined apart
from the regime close to the singularity. Near $y = 0$, all three
solutions (\ref{sol-n}) lead to the following expression for the
bulk potential
\be
V_{B}(\phi) = 3 E - \frac{3}{2}\,c^2 y_1^2\,
\exp\Bigl(\frac{2 \phi}{c |y_{1}|}\Bigr)\,.
\label{potential}
\ee
The scalar field, near the singularity, behaves as: $\phi \simeq 
-c |y_1|\,\ln|y|$, and therefore diverges, as $y \rightarrow 0$, causing
the bulk potential to diverge as well. The sign of $c^2$, appearing in
front of the exponential, determines whether the potential diverges
towards plus or minus infinity. For $c^2>0$, $\phi'^2$ is also positive
and the kinetic term of the scalar field has the correct sign; however,
the potential diverges towards minus infinity, being unbounded from below. 
In this case, the introduction of a second brane to shield the singular
point in the bulk is imperative. If, on the other hand, we consider
the $c^2 < 0$ case, then, we end up with a `tachyonic' kinetic term
for the scalar field; nevertheless, an infinitely-high potential barrier
is rising in front of the singularity in this case, thus shielding the
singular point and allowing, in principle, single brane configurations.
Nevertheless, in what follows, we will introduce a
second brane to shield the singular point even when the latter case
is considered.

The {\it jump} conditions, for the solution presented above, follow from 
Eqs. (\ref{jumpn}) and (\ref{jumpphi}) upon substituting the expressions
for the warp factor (\ref{sol-n}) and the scalar field
(\ref{backreaction3'}) and setting $b=1$ (here, we show the result only
for the case $3 E +
\Lambda_{B} < 0$). They take the form
\ba
V_1 +\Lambda_1 = \frac{6 \omega}{\kappa^2_5}\,\coth(\omega |y_1|)\,,
&~&  \quad 2c=\frac{\partial V_{1}(\phi)}
{\partial \phi} \Biggl|_{y=y_{1}}\,,\label{jump1b}\\[2mm]
V_2 +\Lambda_2 = -\frac{6 \omega}{\kappa^2_5}\,\coth(\omega |y_2|)\,,
&~& \quad 2c=-n_2\,\frac{\partial V_{2}(\phi)}
{\partial \phi} \Biggl|_{y=y_{2}}\,,
\label{jump2b}
\ea
and may lead to the fixing of the location of the two branes in terms of
the remaining fundamental parameters of the theory.

Let us finally check the form of the Friedmann equation on the brane in 
this case. As in the previous subsection, we must first define the
4D gravitational constant. In terms of non-conformal coordinates,
Eq. (\ref{effective4}) still holds, with the factor $n^5$, appearing
in the first line, replaced by $n^4$. Then, we may write
\be \frac{1}{\kappa_{4}^2} \equiv 
\frac{1}{\kappa_{5}^2} \oint dy\,n^2 = 
\frac{1}{\kappa_{5}^2}\,{\cal I}(y_1, y_2)=
\frac{|y_2|-|y_1|}{\kappa^2_5 \sinh^2(\omega y_1)}+ 
\sum_{i=1}^{2} n_{i}^2 \frac{V_{i} + \Lambda_{i}}{6 \omega^2}\,, 
\label{Newton2} \ee 
where we have used again the solution for $\omega^2>0$, and defined
${\cal I}(y_1, y_2)$ as the integral of $n^2$ over the internal compact
space. Turning to the form of the effective cosmological constant on
the brane and using the fact that now $\hat R^{(y)}=-\Bigl(12 n'^2/n^2+
8n''/n\Bigr)$, we find 
\be
\Lambda_{eff} = \frac{3 H^2}{\kappa^2_5}\,{\cal I}(y_1, y_2)\,.
\ee
In the above, we have also used the jump conditions 
(\ref{jump1b})-(\ref{jump2b}). Eliminating the integral 
${\cal I}(y_1, y_2)$ from the above equation by making use of
the definition of the 4D Newton's constant (\ref{Newton2}), we 
recover once again the conventional 4D Friedmann equation
(\ref{Friedmann1}).

\section{Stability analysis}
\setcounter{equation}{0}

In this section, we perform a stability analysis of the solutions found
in the previous section under small, time-dependent perturbations by
using the stabilization constraints that were recently found in \cite{kop}.
Here, we will closely follow the method and notations used in that work
deviating only when the needs of the particular solutions presented
in this article demand it.

\subsection{Stability behaviour of the solutions with $V_B=0$}

We start our analysis by perturbing the metric ansatz (\ref{metric1}),
written in terms of conformal coordinates, in the following way
\be
ds^2 = n^2(t,y)\,\Bigl[ -dt^2 + a^2(t)\,\gamma_{ij}\,dx^i dx^j  
+ b^2(t)\,dy^2 \Bigr]\,.
\label{metrica}
\ee
As in \cite{kop}, we are assume that the time-dependence of the
scale factor $b$ along the extra dimension induces a time-dependence
to the warp factor $n$. Although the method that we follow
is identical to that in \cite{kop}, we need to repeat part of the
calculation: a different metric ansatz is used here, a fact that
changes the final expressions of the stabilization constraints.

Starting from the gravitational part of the action, we need to express the
five-dimensional scalar curvature in terms of the metric function appearing
in (\ref{metrica}). It takes the form 
\be
\hat{R} = \frac{1}{n^2} {\cal R}^{(4)} - \frac{2}{b} D_{\mu}D^{\mu} b +
\hat{R}^{(y)} + {\cal L}_{kin}\,,
\label{hatR}
\ee
where
\be
{\cal R}^{(4)} = \frac{6 \dot{a}^2}{a^2} + \frac{6 \ddot{a}}{a} +
\frac{6 k}{a^2},
\hspace{.5in}
\hat{R}^{(y)} = -\frac{1}{b^2}\,\biggl(\frac{4 n'^{2}}{n^4} +
\frac{8 n''}{n^3}\biggr)\,.
\label{Ry}
\ee
In Eq. (\ref{hatR}), ${\cal L}_{kin}$ contains time-derivatives of $n$ and
$b$ that will eventually give rise to the kinetic term of the canonically
normalized radion field. Here, we are interested in the stability behaviour
of the solutions which can be found from the expression of the effective
potential, therefore, we highlight only the points of our analysis that
lead to the form of this quantity. 

Now, the gravitational effective action can be written as :
\be
S_G = - \int d^4x\,\sqrt{-g_4}\, 
\biggl\{ -A(b)\,\frac{{\cal R}^{(4)}}{2 \kappa_{4}^2} -
\frac{1}{2 \kappa_5^2} \int dy\,n^5 b\,\Bigl({\cal L}_{kin} + \hat{R}^{(y)}
\Bigr)\biggr\}
\label{S-G}
\ee
where $\sqrt{-g_4}=a^3$, and the conformal factor $A(b)$ is :
\be
A(b) = \frac{\kappa_{4}^2}{\kappa_5^2} \oint dy\,b\,n^{3}(t,y)\,.
\label{A}
\ee 
A conformal transformation of the four-dimensional metric, $(g_4)_{\mu\nu}=
(\bar g_4)_{\mu\nu}/A(b)$, removes the coupling between $b$ and
${\cal R}^{(4)}$ and brings the total effective action to the form
\be
S = - \int d^4x\,\sqrt{-\bar g_4}\,\Bigl\{
-\frac{\bar{{\cal R}}^{(4)}}{2 \kappa_{4}^2} +
\bar{{\cal L}}_{kin} + \bar{V}_{eff} \Bigr\}\,,
\label{Smod}
\ee
where the effective potential $\bar{V}_{eff}$ is given by:
\be
\bar{V}_{eff}(b) = \frac{1}{A^2(b)} \oint dy\,n^{5} b\,\Biggl\{ 
-\frac{\hat{R}^{(y)}}{2 \kappa_5^2} + \frac{\phi'^2}{2b^2 n^2} + 
\sum_{i}\,(V_{i} + \Lambda_{i})\,\frac{\delta(y - y_{i})}{b n} \Biggr\}\,.
\label{effectiveV}
\ee
The integral over $y$ is performed over a compact dimension, with $y$
taking values in the `circle' consisting of the symmetric intervals
$(y_1,y_2)$ and $(y_2,y_1)$ while passing across the branes. 

The {\it extremization} and {\it stabilization} constraints follow
from Eq. (\ref{effectiveV}) by taking derivatives with respect to $b$.
We may use the fact that $b$ is always multiplied by $y$ in the
action to define a new coordinate $\xi=b y$. Then, the effective potential
may be rewritten as
\ba
A^2(by_{1}, by_{2})\,\bar{V}_{eff} &=& 2 \int_{by_{1}}^{by_{2}}
d \xi\,n^3\,\biggl( -\frac{6}{\kappa_5^2} \frac{n'^2}{n^2} 
+ \frac{\phi'^2}{2}\biggr)
+ \sum_{i = 1}^2 n^4(by_{i})\,(V_{i} + \Lambda_{i} )\,, 
\label{effectiveV1}
\ea
where now primes denote derivatives with respect to $\xi$. The {\it
extremization} constraint follows by taking the first derivative of
(\ref{effectiveV1}) with respect to $b$. Then, we obtain:
\be
A^2\,\frac{d \bar{V}_{eff}}{d b} =  2 y_{i} n_{i}^2
\biggl( -\frac{6}{\kappa_5^2} \frac{n_{i}'^2}{n_{i}^2} 
+ \frac{\phi_i'^2}{2}\biggr) \Biggl |_{i=1}^{i=2} 
+ \sum_{i = 1}^2 y_{i} \frac{\partial}{\partial (by_{i})} 
[n_{i}^4 ( V_{i} + \Lambda_{i} )] - 2 A \frac{d A}{d b} \bar{V}_{eff}\,. 
\label{extrema0}
\ee

In order to simplify the above expression, we may now use the {\it jump}
conditions in conformal coordinates, that relate $n_i'$ and $\phi_i'$ to
$V_i$'s and which are valid for the static solutions. They can be written
as 
\be
\frac{n'_i}{n_{i}} = \mp \frac{\kappa_5^{2}}{6}\,n_i\,\Bigl 
[\Lambda _{i}+V_{i}(\phi_{i})\Bigr]\,, \qquad 
\phi'_i = \pm \frac{n_i}{2}\,\frac{\partial V_{i}(\phi)}{\partial \phi}
\Biggl|_{y=y_i}\,. 
\label{jump-a}
\ee
for $i=1,2$, respectively. In addition, we may use Eq. (\ref{A}) to derive
the first derivative of $A$ with respect to $b$ 
\be
\frac{d A}{d b} = 2 \frac{\kappa_{4}^2}{\kappa_5^2}\,y_{i} n_{i}^3
\biggl|_{i=1}^{i=2}\,.
\label{dA}
\ee
Finally, we need to know the value of $\bar{V}_{eff}$ evaluated at the
static solution. From the effective action (\ref{Smod}) and for 
${\cal L}_{kin}=0$, we may easily see that $\bar{V}_{eff}$ is the
effective cosmological constant on the brane and it is given by 
\be
\bar{V}_{eff} = \frac{\bar{{\cal R}}^{(4)}}{4 \kappa_{4}^2}  = 
\frac{{{\cal R}}^{(4)}}{4 \kappa_{4}^2 A}=
\frac{3 H^2}{\kappa_{4}^2 A}\,.
\label{Veff}
\ee
Alternatively, the above result may follow by writing $\bar{V}_{eff}=
\Lambda_{eff}/A$ and using Eq. (\ref{Friedmann1}).
If we put everything together, the {\it extremization} constraint becomes
\ba
A^2\,\frac{d \bar{V}_{eff}}{d b} &=&  y_{i}\,n_{i}^5
\Biggl[-\frac{12}{\kappa_5^2} \frac{H^2}{n_{i}^2} +
\frac{\kappa^2_5}{3}\,(\Lambda_i+V_i)^2 -\frac{1}{4}\,\biggl(
\frac{d V_i}{d \phi}\biggr)^2\Biggr] \Biggl |_{i=1}^{i=2} \equiv 0\,. 
\label{extrema}
\ea
The above expression is indeed zero as can be seen by using the explicit
{\it jump} conditions (\ref{jump1})-(\ref{jump2}). Thus, we may
confirm that the solution (\ref{n=b}), together with (\ref{inflation}),
extremizes the radion effective potential. 

The stability behaviour of those solutions under small, time-dependent
perturbations will become manifest in the sign of the second derivative
of the effective potential. The {\it stabilization} constraint, that
requires a positive second derivative, can be obtained by differentiating
Eq. (\ref{extrema0}) one more time with respect to $b$. If we
again use the jump conditions (\ref{jump-a}) as well as the scalar field
equation in the bulk, we find 
\ba
A^2 \frac{d^2 \bar{V}_{eff}}{d b^2} &=& \sum_{i=1}^2 \kappa_5^2 y_{i}^2 
n_{i}^6\,( V_{i} + \Lambda_{i} )\,\Biggl[ 
\frac{5 \kappa_5^2}{18}\,( V_{i} + \Lambda_{i} )^2 - \frac{13}{24} 
\biggl( \frac{d V_{i}}{d \phi} \biggr)^2 \Biggr] \nonumber \\[2mm]
&+& \sum_{i=1}^2 
\frac{1}{4}\,y_{i}^2 n_{i}^6\,\frac{d^2 V_{i}}{d \phi^2} 
\biggl(\frac{d V_{i}}{d \phi}\biggr)^2 
- 2 \Biggl[ \biggl( \frac{d A}{d b} \biggr)^2 + A \frac{d^2 A}{d b^2}
\Biggr] \bar{V}_{eff}\,. 
\label{staba}
\ea
In the above expression, a term proportional to $d \bar V_{eff}/ d b$ has
been already dropped due to the {\it extremization} constraint. If we use
the jump conditions (\ref{jump1})-(\ref{jump2}), the second derivative takes
the form 
\ba 
A^2 \frac{d^2 \bar{V}_{eff}}{d b^2} &=& {\cal X} 
+ \frac{12 H^3}{\kappa_5^2 \sinh(3 H |y_{1}|)} \Biggl\{ - 3 
\frac{[y_{2} \sinh(3 H y_{2}) - y_{1} \sinh(3 H y_{1})]^2}{\cosh(3 H y_{1})
- \cosh(3 H y_{2})} \label{staba1} \\[2mm]
&+& 2 \Bigl[y_{1}^2 \cosh(3 H y_{1}) - y_{2}^2 \cosh(3 H y_{2})\Bigr] -
8 \Bigl[ \frac{y_{1}^2 \cosh(3 H y_{1})}{\sinh^2(3 H y_{1})} - 
\frac{y_{2}^2 \cosh(3 H y_{2})}{\sinh^2(3 H y_{2})} \Bigr] \Biggr\}\,,
\nonumber 
\ea
where
\be
{\cal X} = \sum_{i=1}^2\,\frac{12 H^2}{\kappa^2_5}\,\frac{y_i^2\,n_i^4}
{\sinh^2(3 H y_i)}\,\frac{d^2 V_{i}}{d {\phi}^2}
\label{C2-a}
\ee
is the term involving second derivatives of the interaction terms of the
scalar field on the branes. Let us, for the moment, ignore this term and
concentrate on the expression inside curly brackets in Eq. (\ref{staba1}).
A simple numerical analysis will reveal the sign of this combination. In
Figure 1, we display the value of this expression, as a function of $H$
and the inter-brane distance parameterized by the location of the second
brane, $y_2 \in (y_1,0)$ - for simplicity, we fix $y_1=-1$. As one can
see, this combination is always negative and becomes more negative as both
$H$ and the inter-brane distance increases.  
Eqs. (\ref{staba1})-(\ref{C2-a}) are valid for $H^2>0$, that corresponds to
a de Sitter four-dimensional spacetime. Solutions with a positive cosmological
constant on the brane have been derived in the literature before and they
were shown to be unstable \cite{dS}. However, the presence of a bulk scalar
field, with non-trivial brane interaction terms, may significantly modify
this picture: the destabilizing behaviour displayed in Fig. 1 may be
counterbalanced by the ${\cal X}$-term (\ref{C2-a}) by choosing appropriately
the interaction terms $V_i$. If $d^2 V_i/d \phi^2$ are positive and large
enough, stable brane-world solutions with a positive effective cosmological
constant can easily arise in the framework of this model.

\begin{center}
\begin{figure}[t]
\centerline{\psfig{file=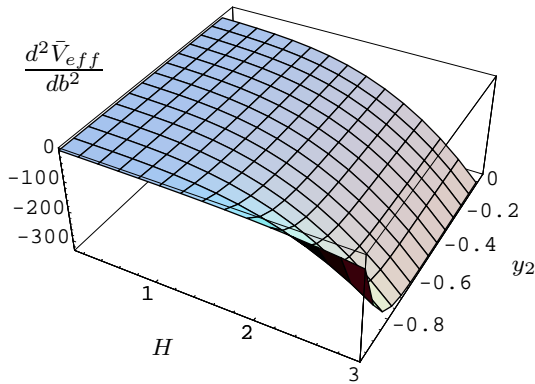, height=5cm}}
\caption{\it The behaviour of the second derivative of the radion
effective potential $\bar V_{eff}$ is displayed as a function of $H$
(for $H^2>0$) and the inter-brane distance.}
\end{figure}
\end{center}
\vspace{-1.5cm}

In the flat-brane limit, $H \rightarrow 0$, the expression inside curly
brackets reduces to zero. This is also in agreement with results in the
literature that consider flat-brane solutions which are saddle points
(flat directions of the potential) \cite{dS, kop}. However, even in the
flat limit, the presence of the scalar field allows for a non-trivial
${\cal X}$-term and the second derivative is written as
\be
A^2 \frac{d^2 \bar V_{eff}}{db^2}= \sum_{i=1}^2\,\frac{4}{3 \kappa^2_5}\,
n_i^4\,\frac{d^2 V_i}{d \phi^2}\,.
\ee
Again, choosing appropriately the interaction terms of the scalar field,
even the flat-brane solutions \`a la Randall-Sundrum can be made stable.

Finally, in the case of Anti de Sitter spacetime, $H^2=-\tilde H^2<0$, the
expression for the second derivative follows from Eq. (\ref{staba1}) after
making the substitutions: $H \rightarrow i \tilde H$ and $\sinh(3 H |y_i|)
\rightarrow i \sin(3 \tilde H |y_i|)$. In that case, the expression inside
curly brackets turns out to be positive-definite revealing the existence
of a stabilizing force acting on the system. This is again in agreement
with results presented in the literature \cite{dS}. The ${\cal X}$-term in
this case has the form
\be
{\cal X} = \sum_{i=1}^2\,\frac{12 \tilde H^2}{\kappa^2_5}\,\frac{y_i^2\,n_i^4}
{\sin^2(3 \tilde H y_i)}\,\frac{d^2 V_{i}}{d {\phi}^2}\,,
\ee
and it may act again as an extra stabilizing force if $d^2 V_i/d \phi^2 >0$.
It is worth noting that, if the above inequality holds, the ${\cal X}$-term
acts as a {\it universal stabilizing force} independently of the value of the
effective cosmological constant on the brane.


\subsection{Stability behaviour of the solutions with $V_B \neq 0$}

We now turn to the stability analysis of the solutions derived in
section 3.2, that are characterized by a non-vanishing bulk potential
$V_B$. Changing to non-conformal coordinates, the perturbed 
five-dimensional line-element may be written as
\be
ds^2= n^2(t,y)\,[\,-dt^2 + a^2(t)\,\gamma_{ij}\,dx^i\,dx^j\,] + 
b^2(t)\,dy^2\,.
\label{metric2}
\ee
The above metric ansatz is similar to that used in Ref. \cite{kop},
therefore, the details of the analysis that lead to the expressions of
the stabilization constraints can be found there. Here, we present only
the main points of our calculation that allow for the generalization of
those results in the case of a non-vanishing effective cosmological
constant. 

For the above ansatz, the scalar curvature may be expressed as in 
Eq. (\ref{hatR}) with the only difference, relevant to our purposes,
being
\be
\hat R^{(y)}=-\frac{1}{b^2}\,\biggl(\frac{12 n'^2}{n^2}+
\frac{8 n''}{n}\biggr)\,.
\ee
Substituting the expression of $\hat R$ in the action and performing
the same conformal transformation, the effective potential finally
takes the form
\be
A^2(by_{1}, by_{2})\,\bar V_{eff}=2 \int_{b y_1}^{b y_2} d\xi\,n^4\,\biggl(
-\frac{6}{\kappa^2_5}\,\frac{n'^2}{n^2} + \frac{\phi'^2}{2}+
\Lambda_B + V_B\biggr) + \sum_{i = 1}^2 n_i^4\,(V_i+\Lambda_i)\,,
\label{effb}
\ee
where now
\be
A(by_{1}, by_{2})=\frac{2 \kappa^2_4}{\kappa^2_5}\,\int_{b y_1}^{b y_2}
d\xi\,n^2(\xi)\,.
\ee

The {\it extremization} constraint demands the first derivative of the
effective potential, with respect to $b$, to vanish when evaluated at
the background solution (\ref{sol-n}). Differentiating once with respect
to $b$ and using the {\it jump} conditions
\be
\frac{n'_i}{n_{i}} = \mp \frac{\kappa_5^{2}}{6}\,\Bigl 
[\Lambda _{i}+V_{i}(\phi_{i})\Bigr]\,, \qquad 
\phi'_i = \pm \frac{1}{2}\,\frac{\partial V_{i}(\phi)}{\partial \phi}
\Biggl|_{y=y_i}\,,
\label{jump-b}
\ee
for $i=1,2$, respectively, we obtain:
\ba
A^2\,\frac{d \bar{V}_{eff}}{d b} =  2 y_{i}\,n_{i}^4
\Biggl[\Lambda_B + V_B +\frac{\kappa^2_5}{6}\,(\Lambda_i+V_i)^2 -
\frac{1}{8}\,\biggl(
\frac{d V_i}{d \phi}\biggr)^2\Biggr] \Biggl |_{i=1}^{i=2} 
-2A\,\frac{d A}{d b}\,\bar{V}_{eff}\Biggl |_{i=1}^{i=2}\,.
\label{extremb}
\ea
The expression of $dA/db$ is given, in this case, by Eq. (\ref{dA}) with
$n_i^3$ replaced by $n_i^2$, while $\bar V_{eff}$ is still given by
Eq. (\ref{Veff}). Finally, we may use the relation
\be
V_{B}(y_i) = 3E -\frac{3}{2}\,\phi'^2_i = 3E - \frac{3}{8}\, 
\biggl( \frac{d V_i}{d\phi} \biggr)^2\,. 
\label{La}
\ee
If we put everything together, the {\it extremization} constraint becomes
\be
A^2 \frac{d \bar{V}_{eff}}{d b} =  y_{i}\,n_{i}^4 \Biggl[
- \frac{12}{\kappa_5^2} \frac{H^2}{n_{i}^2} + \frac{\kappa_5^2}{3} 
( V_{i} + \Lambda_{i} )^2 - \Bigl( \frac{d V_{i}}{d \phi} \Bigr)^2 + 
2(3E + \Lambda_{B})  \Biggr] \Biggr|_{i=1}^{i=2} \equiv 0\,.
\label{extremb2}
\ee
We can easily check that the second equality indeed holds if we use the
{\it jump} conditions (\ref{jump1b})-(\ref{jump2b}).

After differentiating twice Eq. (\ref{effb}) with respect to $b$, we
arrive at  the {\it stabilization} constraint. Once again, using the {\it
jump} conditions (\ref{jump-b}) and the equation of motion of the scalar
field in the bulk (\ref{fieldeqb}), together with Eq. (\ref{case2}), we
arrive at 
\ba
A^2 \frac{d^2 \bar{V}_{eff}}{d b^2} &=& \sum_{i = 1}^2 
y_{i}^2 n_{i}^4 \kappa_5^2\,(V_{i} + \Lambda_{i}) 
\Biggl[ \frac{2 \kappa_5^2}{9}\,( V_{i} + \Lambda_{i})^2
+ \frac{4}{3}\,(3 E + \Lambda_{B}) - \frac{3}{4}\, 
\biggl( \frac{d V_{i}}{d \phi} \biggr)^2 \Biggr] \nonumber \\[2mm]
&+& \frac{1}{4} 
\sum_{i=1}^2 y_{i}^2 n_{i}^4 \,\frac{d^2 V_{i}}{d \phi^2}\, 
\biggl( \frac{d V_{i}}{d \phi} \biggr)^2 
- 2 \Biggl[ \biggl( \frac{d A}{d b} \biggr)^2 + A \frac{d^2 A}{d b^2}
\Biggr] \bar{V}_{eff}\,.
\label{stab3}
\ea

Let us first concentrate on the case with  $\omega^2>0$. By using the
corresponding solution for the warp factor from Eq. (\ref{sol-n}) and 
the {\it jump} conditions (\ref{jump1b})-(\ref{jump2b}), we obtain the
final expression
\be
A^2 \frac{d^2 \bar{V}_{eff}}{d b^2} = {\cal X} + 
\frac{\omega}{\kappa^2_5}\,\Biggl\{
\biggl(\frac{12 \omega^2}{\sinh^2(\omega y_1)}-5 \kappa^2_5 c^2\Bigl)
\,{\cal Y} +
16\,\biggl(\kappa^2_5 c^2-\frac{3 \omega^2}{\sinh^2(\omega y_1)}
\biggl)\,{\cal Z}\Biggr\}\,,
\label{stabw-pos}
\ee
where we have defined ${\cal X}$, ${\cal Y}$ and ${\cal Z}$ as:
\be
{\cal X} = \sum_{i=1}^2 c^2 y_{i}^2 n_{i}^2\,\frac{d^2 V_{i}}{d \phi^2}\,,
\label{calX} 
\ee

\be
{\cal Y}= \frac{ y_{1}^2 \sinh(2 \omega
|y_{1}|) - y_{2}^2 \sinh(2 \omega |y_{2}|)}{\sinh^2(\omega y_1)}\,,
\label{calY}
\ee

\be
{\cal Z} =  \frac{1}{\sinh^2(\omega y_1)}\,
\frac{\Bigl[y_2 \sinh^2(\omega y_2)- y_1 \sinh^2(\omega y_1)\Bigr]^2}
{2\omega (y_1-y_2) + \sinh(2\omega |y_1|)-\sinh(2 \omega |y_2|)}\,.
\label{calZ}
\ee
       	       
In Eq. (\ref{stabw-pos}), the constraint relating $\chi$ and $\omega$
that appears in Eq. (\ref{sol-n}),  may be rewritten as
\be
H^2=\frac{\omega^2}{\sinh^2(\omega y_1)} - \frac{\kappa^2_5 c^2}{3}\,,
\label{Hdef}
\ee
and has been used to eliminate one parameter from the expression of the
second derivative. The value of this quantity still depends on $\omega$
and $c^2$, that parametrize the size of the bulk quantities, $\Lambda_B$
and 
${\cal L}_B(\phi)$, and the inter-brane distance, parametrized by the
values of $y_i$. For simplicity, we may fix the locations of the two
branes at $y_i=-1$ and $y_2=-0.5$, while the singularity lies
at $y=0$. As was pointed out in section 3.2, the parameter $c^2$
may take both positive and negative values; this may lead to different
behaviours of the background solution under time-dependent perturbations,
as we will now see. In Figure 2, we display the value of the second
derivative of the effective potential - the expression inside curly
brackets in Eq. (\ref{stabw-pos}) - for both signs of $c^2$ and we
comment on the results below:

\begin{center}
\begin{figure}[t]
\centerline{\hspace*{-1cm}\hbox{\psfig{file=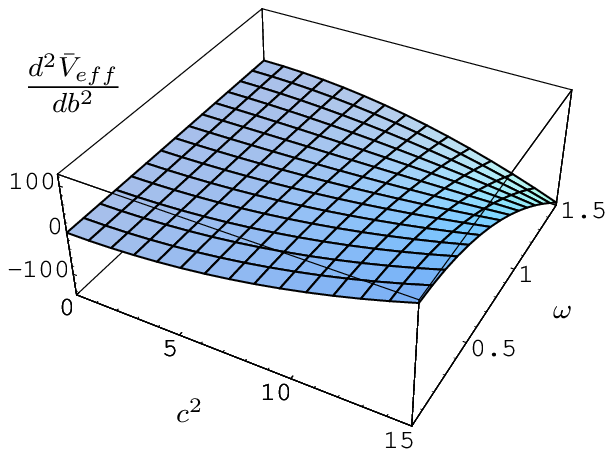, height=5cm}}
\hspace*{0.5cm}
{\psfig{file=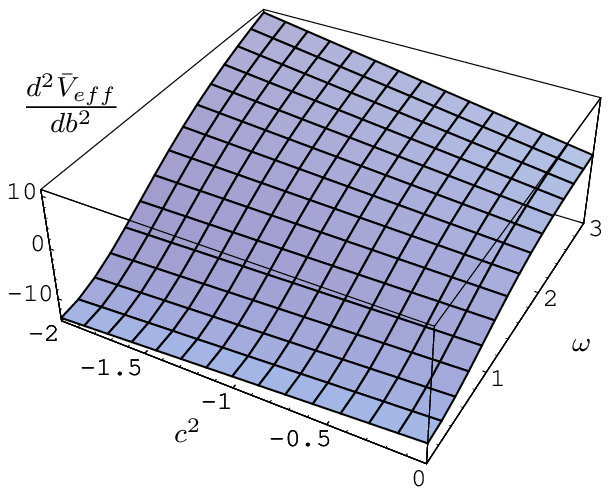, height=5cm}}}
\caption{\it The two plots depict the second derivative of the radion
effective potential $\bar V_{eff}$ as a function of $\omega$ (with
$\omega^2>0$) and for positive and negative values of $c^2$, respectively.}
\end{figure}
\end{center}
\vspace*{-1.3cm}
%
\begin{itemize}
\item \underline{$c^2>0$}: The second derivative assumes positive values
for large enough $c^2$ but small $\omega$. In other words, the background
solution is more stable when the kinetic term of the scalar
field is larger and the bulk cosmological constant is smaller. Eq.
(\ref{Hdef}) then reveals that this particular regime of parameters
corresponds to solutions with a {\it negative} cosmological constant on
the brane. 

\item \underline{$c^2<0$}: Positive values for the second derivative
can be achieved for large enough values of both $c^2$ and $\omega$.
In this case, the effective cosmological constant on the brane is
{\it positive} by definition and therefore physically interesting, stable
solutions may emerge in this case. 
\end{itemize}
It is worth noting that, in the above cases, we obtain positive sign for
the second derivative of the radion potential even if we assume that
$d^2 V_i/d \phi^2 =0$, contrary to what happened in section 4.1.  An
extra stabilizing force may arise, for $c^2>0$ or $c^2<0$, if
$d^2 V_i/d \phi^2 >0$ or $d^2 V_i/d \phi^2 <0$, respectively.
Let us finally note that in the parameter regimes where the solution
is clearly stable or unstable, the sole effect of the inter-brane distance
is to change the magnitude of the radion mass squared: larger
inter-brane distances imply larger (in absolute value) second
derivatives. In the intermediate regime, where the solutions struggle
between stability and instability, stable solutions may arise when
the second brane is located closer to the first one and away from the
singularity.
	       
In the special case of $\omega=0$, the expression of the second derivative
of the effective potential may be obtained from Eq. (\ref{stabw-pos}) by
taking the limit $\omega \rightarrow 0$. Then, we find
\be
A^2 \frac{d^2 \bar{V}_{eff}}{d b^2} = 
\sum_{i=1}^2 \frac{c^2 y_{i}^4}{y_1^2}\,\frac{d^2 V_{i}}{d \phi^2} + 
\frac{2}{\kappa^2_5 y_1^2}\,\Bigl(\kappa_5^2 c^2-{6}{\chi^2}\Bigr)
\,(|y_1|^3-|y_2|^3)\,.
\label{stabw-0}
\ee
The second term has a positive value only if $\kappa^2_5 c^2>6\chi^2 >
0$. In this case, the first term also contributes positively to the
second derivative if
$d^2 V_i/d \phi^2 >0$. In this case, stable solutions with a
negative cosmological constant on the brane arise.

\begin{center}
\begin{figure}[t]
\centerline{\psfig{file=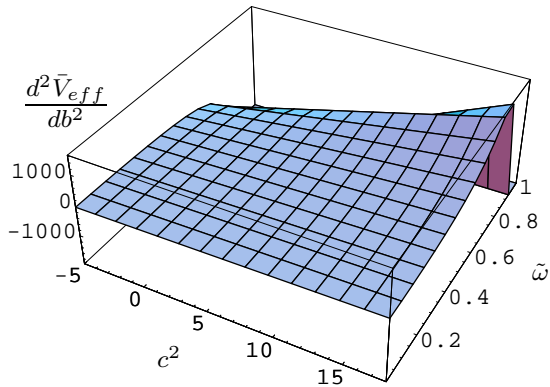, height=5.2cm}}
\caption{ \it The plot depicts the second derivative of the radion effective
potential $\bar V_{eff}$ as a function of the $\tilde \omega$ and $c^2$
parameters.}
\end{figure}
\end{center}
\vspace{-1.4cm}
%

Finally, we turn to the case with $\omega^2<0$. The expression of the
second derivative may again be obtained from Eq. (\ref{stabw-pos}) by
making the substitutions:  $\omega \rightarrow i \tilde \omega$ and
$\sinh(\omega y_i)\rightarrow i \sin(\tilde \omega y_i)$. Given the
new form of the warp factor, the two branes should now be located between
the two singularities at $y=0$ and $\tilde \omega y=-\pi$. We therefore
set $y_1=-\pi$ and we allow $\tilde \omega$ to vary in the interval (0,1)
- $y_2$ is placed in between, with the size of the inter-brane distance
having no effect on the behaviour of the second derivative. In Figure
3, we display the value of the second derivative as a function of $c^2$
and $\tilde \omega$. We may easily see that stable solutions arise for
large enough, positive values of $c^2$, with the solutions becoming
more and more stable as the value of $\tilde\omega$ increases. 
An unstable regime appears when $\tilde \omega \rightarrow 1$, which
is equivalent to placing the first brane very close to the singularity
at $\tilde \omega y=-\pi$. By using Eq. (\ref{Hdef}), we may see that
the stable regime corresponds to solutions with a {\it negative}
cosmological constant on the brane.

\section{Conclusions}
\setcounter{equation}{0}

There has been a lot of work related to the stabilization of the radion 
field that parametrizes the size of the extra dimension in the context of
brane-world models. Most work has focused on the derivation
of solutions with a constant radion field, i.e. a static extra dimension,
with only a few papers investigating whether these solutions correspond
to a true minimum of the radion effective potential. In this paper, we
presented two new, brane-world solutions arising in the presence
of a bulk scalar field, and studied their stability
under time-dependent perturbations of the radion field
demonstrating the existence of phenomeno\-logically interesting,
stable solutions with a positive cosmological constant on the brane.

For our analysis, we used a factorizable ansatz for the line-element
along the brane.
Under the assumption that the total energy of each brane is the sum of
a constant brane tension and the interaction term of a time-independent,
bulk scalar field, the factorization of the 3D scale factor was shown to
be directly related to the stabilization of the extra dimension. 
This relation holds both on and off the brane, as one can see by
using the scale factor {\it jump} conditions and the off-diagonal
component of Einstein's equations, respectively.

The first class of solutions presented here corresponds to a vanishing
bulk potential for the scalar field and a vanishing bulk cosmological
constant. Surprisingly enough, brane-world solutions with non-trivial
warping along the extra dimension did emerge, with the `warping' parameter
being the expansion rate on the brane. These configurations accept a
variety of  time-dependent solutions for the scale factor on the brane,
with flat ($k=0$) or curved ($k=\pm1$) 3D spacetime and a zero, positive
or negative effective cosmological constant. All of the above solutions
are characterized by the presence of a bulk curvature
singularity, which inevitably leads to the introduction
of a second brane in the model. One can show that the {\it jump}
conditions, imposed on the warp factor and the scalar field, lead to
the fixing of the inter-brane distance, in terms of the fundamental
parameters of the theory, and that the conventional Friedmann equation
on the brane is successfully recovered. 

In the second class of solutions that we derived, the presence of a 
non-trivial bulk potential and a bulk cosmological constant was restored.
In this case, the warping of the 5D metric was governed by the bulk
cosmological constant shifted by a constant quantity 
given in terms of the kinetic and potential energy of the bulk
field. The same variety of cosmological solutions on the brane emerge
here as well. The sign of the kinetic term of the scalar field reveals
its nature (normal or tachyonic) and affects the behaviour of the
bulk potential: a normal kinetic term leads to a potential which is
unbounded from below, near the bulk singularity, whereas a tachyonic
kinetic term leads to an infinitely-high potential barrier that may be
used to shield the singularity in single-brane configurations. Here, we
introduced a second brane in order to do so, and we demonstrated that,
as in the first case, the inter-brane distance is fixed and the form of
the Friedmann equation on the brane is recovered.

In the second part of our paper, we investigated the stability of our
solutions under small, time-dependent perturbations of the radion field. 
We derived the {\it extremization} constraints for both types of solutions
and demonstrated that, as expected, they correspond to extrema of the
radion effective potential. The {\it stabilization}
constraints were also derived. These revealed the stability
behaviour of the solutions and the type of extrema to which they
correspond, either minima or maxima. In the first class of solutions,
brane configurations with positive, zero or negative effective
cosmological constant were studied and it was shown that, in the absence
of a scalar field potential, those solutions come out to be local maxima,
saddle points or minima of the radion effective potential, respectively,
in agreement with the literature. However, in our case, the presence of an
extra term, involving second derivatives of the interaction terms of the
scalar field on the branes, acts as a {\it universal} stabilizing force,
independent of the sign of the cosmological constant on the brane, as
long as the second derivatives are positive. In this way, solutions with
a positive cosmological constant on the brane may become stable,
saddle-point solutions \`a la Randall-Sundrum may turn to true minima,
and AdS-type solutions on the brane may be further stabilized. The
results for the second class of solutions found in this paper are even
more interesting: the aforementioned term with the derivatives of the
interaction terms may still act as an extra stabilizing force - upon
appropriate choice of the sign of the second derivatives - nevertheless,
stable solutions can arise even in the case where this term is zero. The
parameter regimes that correspond to stable solutions are determined by
the values of the bulk cosmological constant and the kinetic term of the
scalar field. It is worth noting that the sign of the latter
quantity also defines the sign of the effective cosmological
constant of the stable solution: a normal (positive) kinetic
term gives rise to solutions with negative cosmological constant,
while a tachyonic (negative) one leads to a positive effective
cosmological constant. 

We may, therefore, conclude that the introduction of a bulk scalar
field, in a brane-world model, may successfully lead to a variety of
stable solutions, but more importantly, it may lead to
cosmologically interesting, stable solutions with a positive
effective cosmological constant - a type of solutions that has
been difficult to derive up to now.

\section*{Acknowledgements}
We thank Gregory Gabadadze for  useful discussions.
This work was supported in part by 
DOE grant DE-FG02-94ER40823 at the University of Minnesota.


\end{document}